\documentclass[journal]{IEEEtran}

\usepackage{cite}
\usepackage{epsfig}
\usepackage{graphicx}
\usepackage{amsfonts}
\usepackage{amsmath}
\usepackage{amssymb}
\usepackage{multicol}
\usepackage{psfrag}
\usepackage{color}
\usepackage{subfigure}
\usepackage{stfloats}
\usepackage{footnote}
\usepackage{array}
\usepackage{multirow}
\usepackage{multicol}
 \usepackage{makecell}
\usepackage{algorithm}
\usepackage{algorithmic}
\usepackage{enumitem}

\usepackage[caption=false]{subfig}
\usepackage{comment}
\usepackage{psfrag}
\usepackage{stfloats}

\allowdisplaybreaks

\begin{document}

\bstctlcite{IEEEexample:BSTcontrol}

\title{\huge Terahertz Communications for 6G and Beyond Wireless Networks: Challenges, Key Advancements, and Opportunities}

\author{Akram Shafie,~\IEEEmembership{Graduate Student~Member,~IEEE,} Nan Yang,~\IEEEmembership{Senior~Member,~IEEE,} Chong Han,~\IEEEmembership{Member,~IEEE,} Josep Miquel Jornet,~\IEEEmembership{Senior~Member,~IEEE,} Markku Juntti,~\IEEEmembership{Fellow,~IEEE}, and Thomas K\"urner,~\IEEEmembership{Fellow,~IEEE}

}

\maketitle

\begin{abstract}
The unprecedented increase in wireless data traffic, predicted to occur within the next decade, is motivating academia and industries to look beyond contemporary wireless standards and conceptualize the sixth-generation (6G) wireless networks. Among various promising solutions, terahertz (THz) communications (THzCom) is recognized as a highly promising technology for the 6G and beyond era, due to its unique potential to support terabit-per-second transmission in emerging applications. This article delves into key areas for developing end-to-end THzCom systems, focusing on physical, link, and network layers. Specifically, we discuss the areas of THz spectrum management, THz antennas and beamforming, and the integration of other 6G-enabling technologies for THzCom. For each area, we identify the challenges imposed by the unique properties of the THz band. We then present main advancements and outline perspective research directions in each area to stimulate future research efforts for realizing THzCom in 6G and beyond wireless networks.
\end{abstract}

\section{Introduction}

We live in an era of remarkable expansion of wireless data traffic.  Although the recently launched fifth-generation (5G) networks can offer significant advancements compared to 4G Long-Term Evolution, such networks are inherently limited in supporting the data and connection demands in 2030s, since these demands go far beyond the initial premise of 5G networks. Specifically, the International Telecommunication Union (ITU) predicts that by 2030, the global wireless data traffic will reach an astonishingly 5 zettabytes per month and the number of connected devices might surge beyond 50 billions. This has prompted industries and academia to look beyond 5G and conceptualize the sixth-generation (6G) networks \cite{2020_Mag6G_WalidSaad}.

Among various cutting-edge and promising solutions, terahertz (THz) communications (THzCom), which refers to the communications in the ultra-wide THz band ranging from 0.1 to 10 THz, is envisioned as a highly promising technology for 6G and beyond era \cite{2020_IEEENEtworks}. In particular, the huge available bandwidths in the order of tens up to a hundred gigahertz (GHz) and extremely short wavelengths offer enormous potentials to alleviate the spectrum scarcity and break the capacity limitation of 5G networks, thereby enabling emerging applications that demand an explosive amount of data, such as holographic telepresence, extended reality, and ultra high-speed wireless backhaul. 
Thus, similar to that millimeter wave (mmWave) communications was first envisioned and then utilized for 5G, THzCom will inevitably play an indispensable role in developing 6G and beyond wireless networks~\cite{2018MagCombatDist}.

\textcolor{black}{Despite the prospects, THz band, especially the carrier frequencies above 300 GHz, brings new challenges that have never been addressed at sub-6 GHz and even mmWave frequencies \cite{2019_OJCS_Survey_SalmanRef1}. Particularly,
in addition to the severe spreading loss and high channel sparsity, the THz band is characterized by the unique molecular absorption loss which is frequency and distance dependant. Moreover, the very short wavelength of THz signals makes THz signal propagation highly vulnerable to blockages, since objects with small dimensions can act as impenetrable blockers at THz frequencies. Furthermore, high reflection and scattering losses attenuate the non-line-of-sight (nLoS) rays significantly, triggering the need for line-of-sight (LoS) links for reliable transmission.
All these phenomena lead to distinctive propagation characteristics at the THz band, which necessitates the design and development of new communications and signal processing mechanisms for THzCom.}

\begin{figure*}[!t]
\centering
\includegraphics[width=2.0\columnwidth]{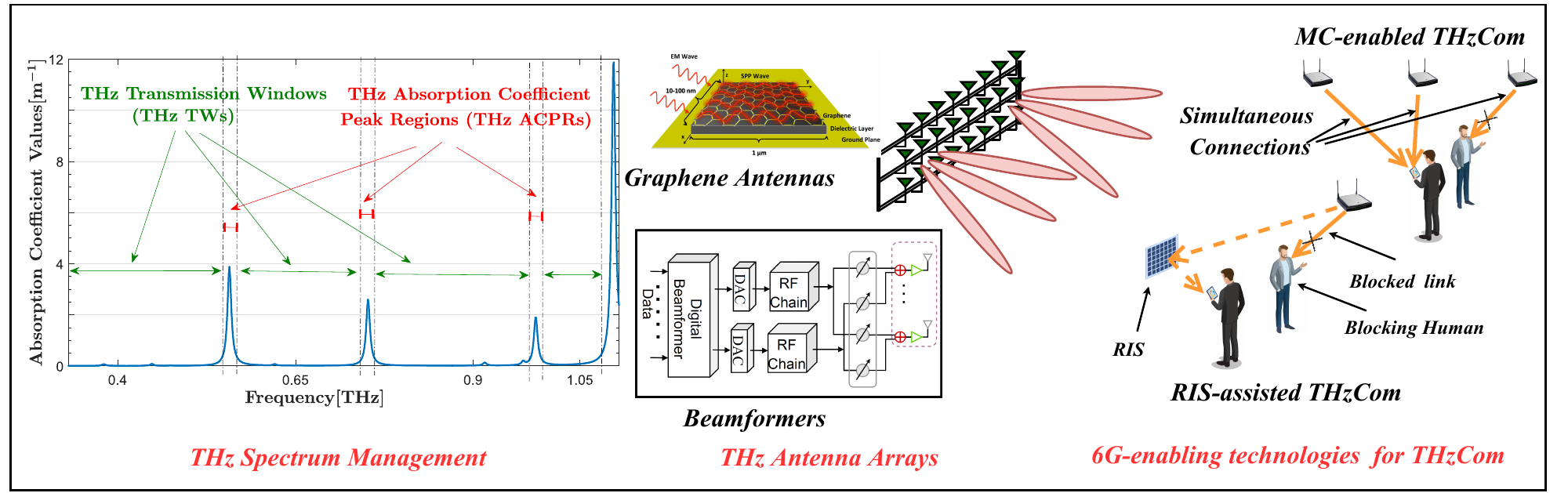}
\caption{Illustration of the three pivotal areas for the development of end-to-end THzCom systems.}\label{Fig:Summary}
\vspace{-2mm}
\end{figure*}

Following the recent breakthroughs on THz hardware technologies, the research on THzCom has accelerated rapidly in recent years and primarily focused on channel measurement and modeling, device technologies, experimental results, and standardization, e.g., \cite{2020_WCM_THzMag_Standardization,2019_OJCS_Survey_SalmanRef1}. Despite such research has uniquely advanced our understanding of THz technologies, the research on end-to-end THzCom systems is still limited.
\textcolor{black}{For example, a recent study \cite{2020_WCM_THzMag_TerahertzNetworks} examined the end-to-end THzCom systems, with a focus on the network layer challenges. Differently, this article provides a holistic and multidimensional view on the challenges, key advancements, and opportunities of end-to-end THzCom systems in the 6G and beyond era, with the primary focus on physical, link, and network layers.}
To this end, we examine three pivotal areas of THzCom which will pave the way for implementing end-to-end THzCom systems in the 6G and beyond era. Specifically, as depicted in Fig.~\ref{Fig:Summary}, we discuss:
\begin{itemize}
\item \textbf{THz band spectrum management} to unlock the inherent potentials offered by the huge bandwidth at the THz band.
\item \textbf{THz band antennas and beamforming} to reap the benefits of processing THz signals which have small wavelengths.
\item \textbf{Joint integration of four 6G-enabling technologies for THzCom} to materialize end-to-end THzCom systems for achieving various QoS targets anticipated by 6G and beyond networks.
\end{itemize}
\textcolor{black}{The first two key areas explore the inherent benefits of the THz band and the third key area discusses the integration of four 6G-enabling technologies into THzCom.}
%
We note that most of previous relevant studies have discussed THzCom systems operating at the sub-THz band, i.e., the carrier frequency from 100 GHz to 300 GHz, with bandwidth of up to several tens of GHz, e.g., the first THzCom standard targeting 6G -- IEEE 802.15.3~\cite{2020_WCM_THzMag_Standardization}. Differently, in this article, we consider a much broader scope by discussing the effects relevant to THzCom systems operating at carrier frequencies above 300 GHz with much larger bandwidths.
\textcolor{black}{A brief summary of the challenges, key advancements, and opportunities discussed in this paper are presented in Table I.}

\section{THz Band Spectrum Management}

The data rate demands of 6G networks are anticipated to be much higher than what the state-of-the-art wireless systems can offer, e.g., 6G networks would demand the peak data rate of up to one terabit per second  (Tb/s), which is 100 times that of 5G \cite{2020_Mag6G_Marco_UseCasesandTechnologies}. To meet this demand, the enormous potential offered by the huge available bandwidths of the THz band must be explored. The frequency-dependent nature of absorption loss divides the entire THz band into ultra-wideband transmission windows (TWs) and absorption coefficient peak regions (ACPRs), as shown in the left of Fig.~\ref{Fig:Summary}. We next discuss the challenges brought by utilizing the spectra in TWs and ACPRs, and present promising future research directions. 

\begin{table*}[!t]
\textcolor{black}{\caption{Summary of the challenges, key advancements, and opportunities discussed in this paper.}\label{tab:Bandwidth}\vspace{-4mm}
\begin{center}
\begin{tabular}{|p{0.015\linewidth}|p{0.0375\linewidth}|p{0.25\linewidth}|p{0.25\linewidth}|p{0.325\linewidth}|}\hline
\multicolumn{2}{|c|}{\textbf{Key Area}}&\textbf{Challenges}& \textbf{Key Advancements}& \textbf{Opportunities/ Future Research Directions} \\ \hline
\multicolumn{2}{|c|}{\rotatebox[origin=r]{90}{\makecell{\textit{THz band~~~~~}   \\ \textit{Spectrum Management~~~~~} }  }} & \vspace{-2.5mm}
\begin{itemize}[leftmargin=*]\renewcommand\labelitemi{$\diamond$}
\item Frequency mapping
\item Distance and frequency dependency of molecular absorption loss
\item Inter-band interference (IBI)
\item Temporal broadening effect
\item Impact of the pressure, temperature, and relative humidity condition of propagation environment\vspace{-3mm}
\end{itemize}
& \vspace{-2.5mm}
\begin{itemize}[leftmargin=*]\renewcommand\labelitemi{$\diamond$}
\item Single-band pulse-based spectrum allocation
\item Multi-band-based spectrum allocation
\item Distance-aware multi-carrier (DAMC) based spectrum allocation
\item Hierarchical bandwidth modulation (HBM)
\item Distance-adaptive absorption peak modulation (DA-APM)
\end{itemize}
& \vspace{-2.5mm}
\begin{itemize}[leftmargin=*]\renewcommand\labelitemi{$\diamond$}
\item Design of power amplifiers and transceivers that are capable of handling multi-GHz bandwidths
\item IBI modelling, analyzing, suppression, and elimination
\item Multi-band-based spectrum allocation with adaptive sub-band bandwidth (ASB)
\item Optimally avoiding spectra at the edges of transmission windows (TWs)
\item HBM when (i) the available bandwidths of different links are similar and (ii) the number of transmission links are higher than two
\item Secure communications exploiting the temporal broadening effect
\item Dynamic spectrum allocation  \vspace{-2.5mm}
\end{itemize}\\\hline
\multicolumn{2}{|c|}{\rotatebox[origin=r]{90}{\makecell{\textit{THz band~~}   \\ \textit{Antennas and}  \\ \textit{Beamforming} }  }}    & \vspace{-2.5mm}
\begin{itemize}[leftmargin=*]\renewcommand\labelitemi{$\diamond$}
\item Device technology and circuit design
\item Low multiplexing gain
\item Channel squit effect
\item Sparse channel
\item Poor multiplexing gain
\item Dynamic channel  \vspace{-2.5mm}
\end{itemize} & \vspace{-2.5mm}
\begin{itemize}[leftmargin=*]\renewcommand\labelitemi{$\diamond$}
\item Frequency up-converting systems
\item Spatial multiplexing designs
\item Widely-spaced hybrid beamforming architectures (HBAs)
\item Dynamic array-of-subarray (DAoS) HBAs \vspace{-2.5mm}
\end{itemize}
& \vspace{-2.5mm}
\begin{itemize}[leftmargin=*]\renewcommand\labelitemi{$\diamond$}
\item Graphene technology
\item Low-cost and energy-efficient nano-transceiver design
\item Channel models with spherical wave propagation
\item THz true-time delays (TTDs)
\item Unified HBAs \vspace{-2.5mm}
\end{itemize}
\\\hline
\multirow{4}{*}{\rotatebox[origin=c]{90}{\textit{6G-enabling Technologies for THzCom}~~~~~~~~~~~~~ }}&\rotatebox[origin=r]{90}{\makecell{\textit{Out-of-band~~~}   \\ \textit{Channel~~~}  \\ \textit{Estimation~~~}}}   & \vspace{-2.5mm}
\begin{itemize}[leftmargin=*]\renewcommand\labelitemi{$\diamond$}
\item Accurate channel estimation
\item Mismatch in the angles-of arrival and the angle spread between lower frequencies and THz signals
\end{itemize} & \vspace{-2.5mm}
\begin{itemize}[leftmargin=*]\renewcommand\labelitemi{$\diamond$}
\item Out-of-band for millimeter wave (mmWave) communications, i.e., estimating the mmWave channels using sub-6G channel measurements\vspace{-3mm}
\end{itemize}
& \vspace{-2.5mm}
\begin{itemize}[leftmargin=*]\renewcommand\labelitemi{$\diamond$}
\item Exploit the structure of the spatial correlation matrix along with interpolation/extrapolation techniques to determine the THz band correlation matrix
\item Exploit the theoretical expressions for the THz band correlation in conjunction with the measured lower frequency correlation to determine the THz band correlation matrix
\vspace{-3mm}
\end{itemize}
\\\cline{2-5}
&\rotatebox[origin=r]{90}{\makecell{\textit{Multi-}   \\ \textit{Connectivity}   \\ \textit{(MC)}}}  & \vspace{-2.5mm}
\begin{itemize}[leftmargin=*]\renewcommand\labelitemi{$\diamond$}
\item Reliability degradation caused by blockages
\item Cell-free architecture with centralized controlling, data processing,
and combining strategies.
\item Increase in interference \vspace{-2.5mm}
\end{itemize} & \vspace{-2.5mm}
\begin{itemize}[leftmargin=*]\renewcommand\labelitemi{$\diamond$}
\item Theoretical evaluation of the benefits of MC\vspace{-3mm}
\end{itemize}
& \vspace{-2.5mm}
\begin{itemize}[leftmargin=*]\renewcommand\labelitemi{$\diamond$}
\item Jointly deployment of MC and HBAs
\item Dynamic AP selection strategies
\item ``Lightweight'' solutions, such as the reactive MC strategy
\item Interference management and mitigation techniques
\item Promising packet scheduling methods for latency reduction, e.g., packet duplication and splitting
\vspace{-3mm}
\end{itemize}
\\\cline{2-5}
&\rotatebox[origin=r]{90}{\textit{RIS~~~~}}   & \vspace{-2.5mm}
\begin{itemize}[leftmargin=*]\renewcommand\labelitemi{$\diamond$}
\item Reliability degradation caused by blockages
\item Reflector array material selections
\item RIS channel characterization
\end{itemize} & \vspace{-2.5mm}
\begin{itemize}[leftmargin=*]\renewcommand\labelitemi{$\diamond$}
\item Design of RIS-aided THzCom systems\vspace{-3mm}
\end{itemize}
& \vspace{-2.5mm}
\begin{itemize}[leftmargin=*]\renewcommand\labelitemi{$\diamond$}
\item Use of both passive and active RIS
\item Form virtual MC links with RIS
\item Graphene for RIS
\item Experimental validation of THz band RIS channels
\vspace{-3mm}
\end{itemize}
\\\cline{2-5}
&\rotatebox[origin=r]{90}{\makecell{\textit{Machine}   \\ \textit{Learning}   \\ \textit{(ML)}}} & \vspace{-2.5mm}
\begin{itemize}[leftmargin=*]\renewcommand\labelitemi{$\diamond$}
\item Near-real-time operation requirement
\item Computational complexity
\item Data acquisition for the ML-training process
\end{itemize} & \vspace{-2.5mm}
\begin{itemize}[leftmargin=*]\renewcommand\labelitemi{$\diamond$}
\item Development of different ML approaches for wireless
communications systems\vspace{-3mm}
\end{itemize}
& \vspace{-2.5mm}
\begin{itemize}[leftmargin=*]\renewcommand\labelitemi{$\diamond$}
\item Techniques to process highly non-stationary THz band training data
\item Efficiently distribute intelligence in the network
\vspace{-3mm}
\end{itemize}
\\\hline
\end{tabular}
\end{center}\vspace{-4mm}}
\end{table*}

\subsection{THz Transmission Windows}

For very short transmission distances (less than one meter), the absorption loss is very low and remains almost the same within the whole THz band. Due to this phenomenon, single-band pulse-based spectrum allocation schemes, where the whole THz band is considered as a single band, as depicted in case (i) of Fig.~\ref{Fig:THzBand}, can be used for applications with very short transmission distances, e.g., nanonetworks and device-to-device communications \cite{HBM1}. Although such schemes can theoretically achieve Tb/s data rates using one-hundred-femto second-long pulses, it is challenging to design low-cost and energy-efficient nano-transceivers that can support smart physical- and link-layer solutions.

When transmission distances are relatively long, the molecular absorption varies significantly within the THz band, making it impractical to use single-band pulse-based spectrum allocation schemes. This requires us to explore efficient spectral allocation strategies, based on carrier-based modulation schemes, for utilizing the bandwidths available within TWs towards micro- and macro-scale THzCom applications. In THz band carrier-based systems, the to-be-transmitted data is first modulated into single-carrier waveforms and/or pulses and then the modulated signals are upshifted to higher THz band frequencies via cascaded frequency multipliers. We next discuss two carrier-based spectrum allocation schemes for THzCom systems.

\subsubsection{Multi-band-based Spectrum Allocation Scheme}

\textcolor{black}{In this scheme, as depicted in case (iii) of Fig.~\ref{Fig:THzBand}, the TWs are divided into a set of non-overlapping sub-bands whose bandwidths are reasonably low as compared to the bandwidths of TWs.}
Then, these sub-bands are exploited to serve a single user or multiple users that simultaneously demand ultra-high data rates. By doing so, the multi-band-based scheme achieves high spectrum efficiency (SE) and low complexity, especially when the absorption loss varies significantly within TWs.

The inter-band interference (IBI) is an important performance limiting factor in the multi-band-based spectrum allocation scheme, especially when each user demands multiple sub-bands to satisfy its QoS requirements. The IBI at the THz band is caused by the power leakage from adjacent sub-bands, due to harmonics generated in frequency multiplying chains during frequency upshifting. Given the high bandwidth observed at the THz band, sub-bands can be well spaced to avoid the effect of IBI. Hence, THzCom systems adopting well-spaced sub-bands can achieve much higher data rates than those envisioned by the state-of-the-art mmWave systems. Despite so, the need to fully utilize the bandwidths available in TWs may arise, in order to satisfy the QoS requirements of 6G and beyond networks. Due to this, when sub-bands are close to each other, the impact of IBI needs to be precisely modeled, analyzed, and then suppressed or eliminated to reap the benefits of multi-band-based schemes. It is noted that the Gaussian approximation has been widely used to model IBI at sub-6 GHz and mmWave frequencies, assuming that neighboring sub-bands have similar channel quality \cite{HBM2}. However, the applicability of this approximation at the THz band needs to be re-examined, since the adjacent THz sub-bands exhibit different channel qualities, due to absorption loss. As for IBI suppression, time-domain front-end filtering, subcarrier weighting, windowing, and orthogonal precoding methods are worth careful investigation. Furthermore, owing to the narrow beams observed at the THz band, beam management solutions can be developed to overcome IBI in multiuser THzCom systems such that the adjacent bands are assigned to transmission links that are located far apart.

\begin{figure}[!t]
\centering
\includegraphics[width=0.95\columnwidth]{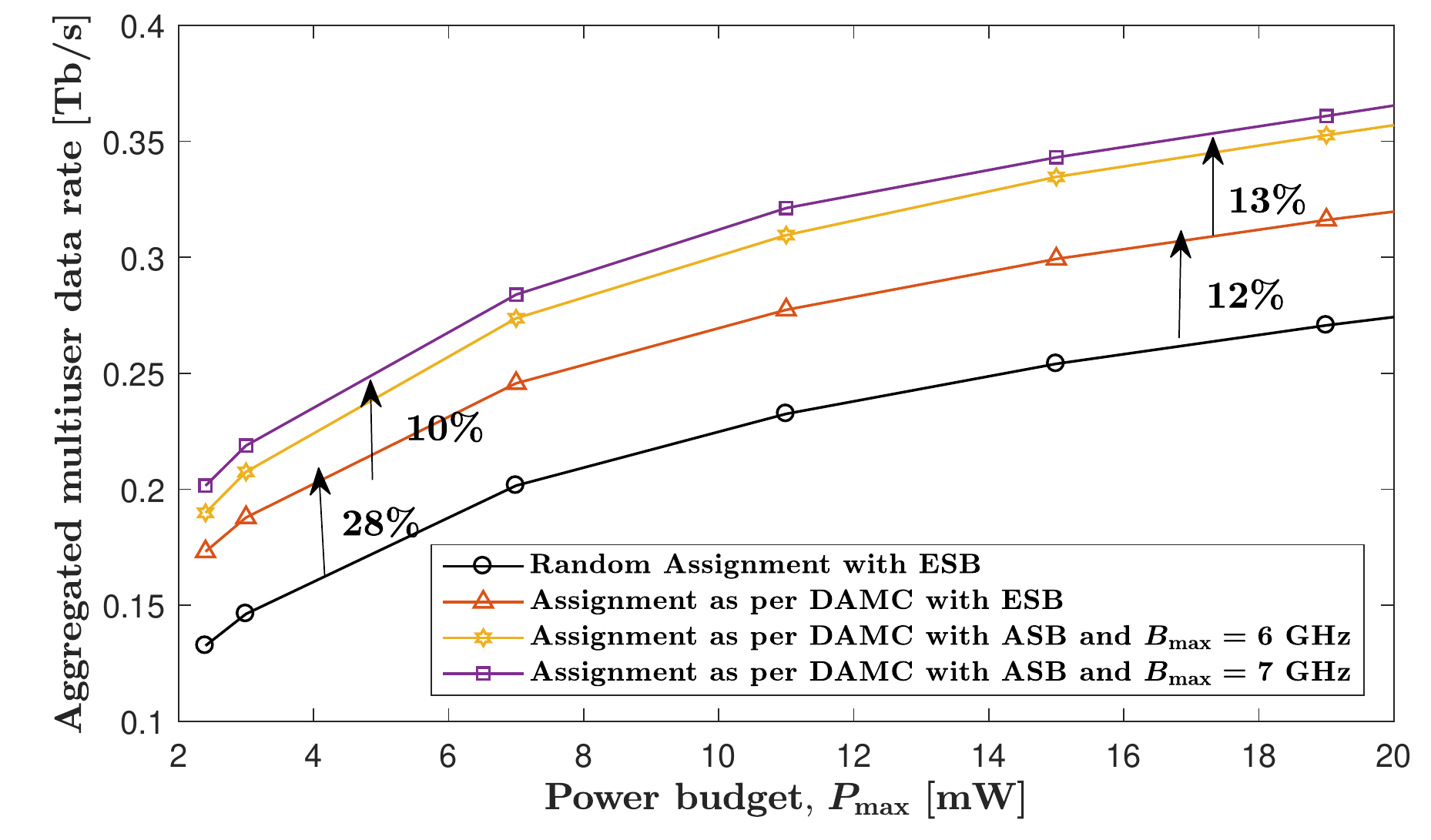} \vspace{-2mm}
\caption{Aggregated multiuser data rate versus the power budget, where $B_{\textrm{max}}$ is the upper bound on the sub-band bandwidth. An indoor environment of size $20~\textrm{m}\times 20~\textrm{m}$ is considered, where the AP and users are equipped with directional antennas with $25~\textrm{dBi}$ gains.}\label{Fig:Num1}
\vspace{-4mm}
\end{figure}

When considering multi-band-based spectrum allocation in multiuser THzCom systems, novel challenges are caused by the distance-dependent nature of absorption loss. To address such challenges, distance-aware multi-carrier (DAMC) based spectrum allocation schemes can be utilized to improve the throughput fairness among users in multiuser THzCom systems \cite{HBM2}. In DAMC-based spectrum allocation schemes, the sub-bands existing in the center region of the TW are assigned to the links with longer distances, while the sub-bands existing at the edges of the TW are assigned to the links with shorter distances. \textcolor{black}{We note that the state-of-the-art DAMC-based multi-band spectrum allocation schemes have only considered equal sub-band bandwidth (ESB), where the TW is divided into sub-bands with equal bandwidth. However, it would be beneficial to explore novel multi-band spectrum allocation schemes with adaptive sub-band bandwidth (ASB) to improve the SE, by dividing the TW into sub-bands with unequal bandwidths.} Moreover, since absorption loss is very high at edge sub-bands, it would be highly beneficial if some spectra at the edges of TWs are not used during spectrum allocation. \textcolor{black}{To illustrate the advantages of the DAMC principle and ASB in multiuser THzCom systems, we use Monte Carlo simulations to plot the aggregated multiuser data rate in Fig. \ref{Fig:Num1}.}
 In this figure, we consider a three-dimensional (3D) THzCom system where an access point (AP) at the center of the indoor environment serves 10 users using 50 GHz spectrum in the first TW above 1 THz. As shown in this figure, DAMC-based spectrum allocation schemes achieve a profound data rate improvement compared to random spectrum allocation schemes. Also, the use of ASB brings noticeable data rate improvement, compared to the use of ESB.

\subsubsection{Multi-TW-based Spectrum Allocation Scheme}

\begin{figure}[!t]
\centering
\includegraphics[width=0.95\columnwidth]{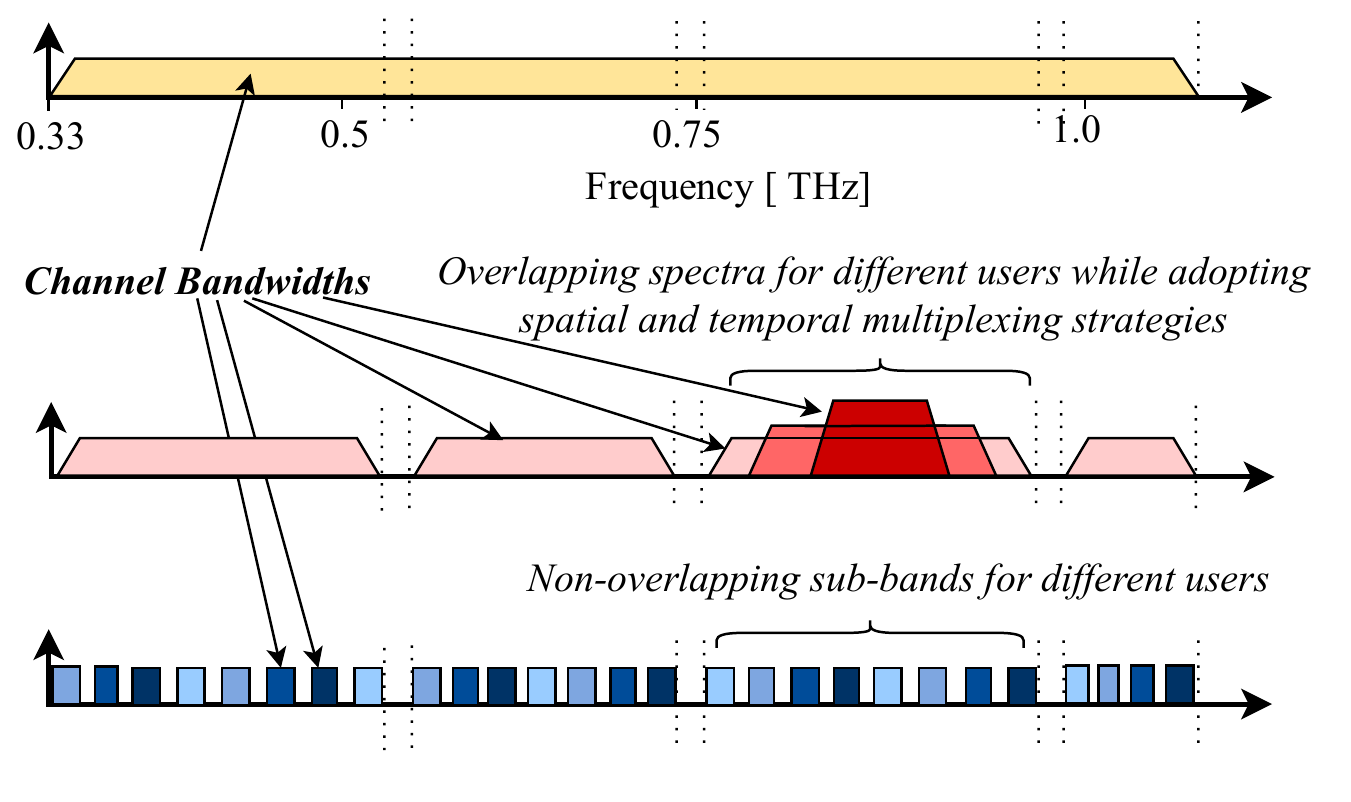}
\vspace{-5mm}
\caption{Illustration of bandwidth allocation for Case (i) single-band pulse-based schemes (top), Case (ii) multi-TW-based schemes (middle), and Case (iii) multi-band-based schemes (bottom).}\label{Fig:THzBand}\vspace{-2mm}
\end{figure}

In this scheme, individual TWs are fully allocated to separate high-speed communications links while exploring wideband signals with bandwidths equal to those of TWs, as shown in case (ii) of Fig.~\ref{Fig:THzBand} \cite{HBM1}.
Open problems related to this scheme include: \textcolor{black}{Determining the optimal usable bandwidth} and modulation order that guarantee very high throughput while ensuring the minimal impairment caused by the distance-dependent nature of absorption loss within TWs, and designing power amplifiers and transceivers that are capable of handling multi-GHz bandwidths.

\textcolor{black}{When this scheme is adopted in multiuser THzCom systems, the same spectrum needs to be shared among multiple users, as shown in case (ii) of Fig.~\ref{Fig:THzBand}, since the number of available TWs within the entire THz band is limited. This necessitates the exploration of spatial and temporal multiplexing strategies. Motivated by this, hierarchical bandwidth modulation (HBM) can be explored, where multiple data streams are transmitted over links with different distances at the same time and within the same TW, by simultaneously adapting both symbol duration and modulation order \cite{HBM1}. Notably, the symbol duration for the links with shorter distances and larger available bandwidths needs to be smaller than that for the links with longer distances.}

\subsection{THz Absorption Coefficient Peak Regions}

Conventionally, ACPRs are not favorable for applications that demand ultra-high data rates, due to the high absorption loss experienced at the spectra in ACPRs. However, the highly exponential nature of absorption loss with transmission distance in ACPRs can be exploited to materialize secure short range THzCom links. Recently, the distance-adaptive absorption peak modulation (DA-APM) has been proposed to enhance covertness when the illegitimate user is located further than the legitimate user, by dynamically modulating signals at ACPRs \cite{2020_Chong_TWC_DistanceAdaptiveAbsorptionPeakModulation}. Inspired by this, open problems related to DA-APM include: Selection of appropriate operating frequency and bandwidth within the ACPR based on the distance of the legitimate link and the required level of secrecy performance, signal design which guarantees the smoothness of its power spectral density to avoid local power leakage and enhance security, and the determination of the optimal transmit beamwidth such that a compromise between secrecy and antenna directional gains is achieved to compensate for path loss.

Apart from DA-APM, the impact of temporal broadening within ACPRs can also be exploited for secure communications, when the illegitimate user is located further than the legitimate user. It is noted that at the THz band, temporal broadening, i.e., pulse broadening in the time domain, is mainly caused by the frequency selective nature of molecular absorption loss that is significant within ACPRs. Specifically, since temporal broadening increases exponentially with distance and the molecular absorption coefficient, we need to determine appropriate operating frequency, bandwidth, and time interval between two consecutive pulses, based on the distance of the legitimate link. This ensures that non-overlapping broadened pulses are detected at the legitimate user, but overlapping broadened pulses are detected at the illegitimate user, which enhances the secrecy performance.

\section{THz Band Antennas and Beamforming}

The very short wavelength of THz signals enables the integration of a huge number of antennas  into transceivers in a very small footprint to form ultra-massive multiple-input multiple-output (UM-MIMO) antennas \cite{2020_Jornet_JSAC_AntennaArray}. It is important to note that such antenna arrays with carefully designed signal processing algorithms can deliver extremely high directional gains, e.g., 30-50 dBi \cite{2018MagCombatDist}; thus, they can be utilized to compensate for the severe spreading and absorption losses experienced at THz band. Moreover, multi-beam architectures can be leveraged for massive user multiplexing in the short range. In the following, we discuss the challenges related to device technologies and a pivotal signal processing algorithm in THzCom systems, namely, beamforming, and present promising future research directions. 

\subsection{Device Technology}

For many years, the challenges associated with on-chip generation, modulation and radiation of THz signals has limited the utilization of the THz band. However, the THz technology gap is progressively being closed, thanks to the major advancements across different device technologies \cite{2020_Jornet_JSAC_AntennaArray}. Today, the majority of THzCom systems rely on frequency up-converting systems.
Different processes can be utilized to build such up-converters, ranging from silicon complementary metal-oxide-semiconductor (CMOS) 
at frequencies of up to 200-300 GHz to the III-V semiconductor-based Hight Electron Mobility Transistor (HEMT) or the Schottky diode technology 
at frequencies above 1 THz. For now, most of the THz testbeds are single-input single-output and rely on external high-gain directional antennas (e.g., horn and dish antennas). Only recently, fully digital antenna arrays with 8 to 16 channels have been demonstrated at frequencies between 100 and 200 GHz \cite{2020_Jornet_JSAC_AntennaArray}.


Beyond well-established technologies, innovative solutions are possible when embracing new materials that open doors to new physics. \textcolor{black}{Inspired by this, the use of graphene has been proposed to develop signal sources, modulators, antennas and antenna arrays, due to its ability to support the propagation of surface plasmon polariton (SPP) waves, theoretically for frequencies as low as 100 GHz~\cite{2020_Jornet_JSAC_AntennaArray}. However, for graphene-based plasmonic devices to operate efficiently in the lower THz band, high electron relaxation times are required. Although theoretically possible, it remains experimentally very challenging; thus, at present, the majority of graphene-based plasmonic devices for THz applications are found at frequencies of 1 THz and beyond~\cite{2020_Jornet_JSAC_AntennaArray}. Interestingly, this is also the frequency range where the conventional electronic approaches to THz wave generation decay, making graphene a promising candidate.}

\subsection{Beamforming}

To materialize the high directional gains promised by THz antenna arrays, efficient beamforming architectures that achieve high SE and energy efficiency EE are desired. Among different beamforming architectures, hybrid beamforming architectures (HBAs) are envisioned for THzCom systems due to their low hardware complexity and capability of achieving simultaneously high SE and EE \cite{2018MagCombatDist,2020_Chong_DAoSA}. Although HBAs have been studied at lower frequencies, the applicability of those HBAs for THzCom systems is arduous since the characteristics of THz channels impose unique challenges. Specifically, high THz channel sparsity leads to poor multiplexing gain, even if a significantly higher number of antennas that are critically spaced (spacing of half of the wavelength between antennas) are used in antenna arrays. Moreover, channel squint effects, i.e., phase shift errors, are introduced during beamforming at the THz band since THzCom systems generally have large bandwidth. Furthermore, the high susceptibility of THz signals towards blockages leads to dynamic channel requirements, resulting in very low efficiency when traditional static HBAs are adopted. These challenges require new HBAs that are different from those designed at lower frequencies.

\begin{figure}[!t]
\centering
\includegraphics[width=0.95\columnwidth]{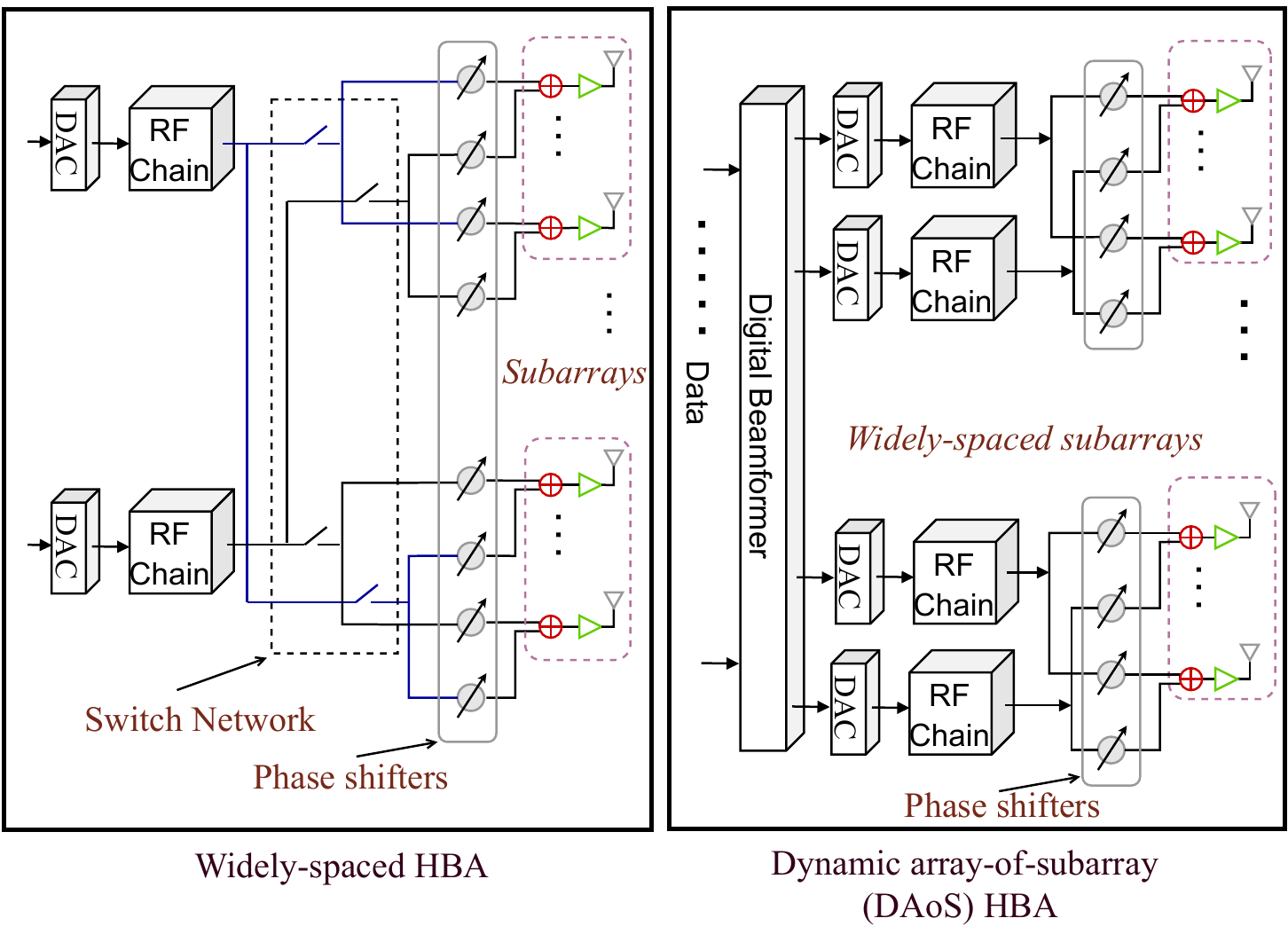}
\vspace{-3mm}
\caption{Illustration of potential hybrid beamforming architectures (HBAs) for THz band antennas.}\label{Fig:HBF}\vspace{-2mm}
\end{figure}


Spatial multiplexing designs, such as the are certainly worthwhile to explore for overcoming the poor multiplexing gain caused by channel sparsity. One such design consists of grouping the antennas into several subarrays and separating these subarrays over hundreds of wavelengths, i.e., widely-spaced, as depicted in Fig.~\ref{Fig:HBF}, aiming to reduce the correlation among these subarrays.    \textcolor{black}{Moreover, widely-spaced HBAs can also act as frequency diverse arrays (FDAs), which can intentionally introduce frequency offsets across array antennas for decoupling the highly correlated channels of legitimate users and illegitimate users, thereby enhancing the security of THzCom systems.} Note that the same radio frequency (RF) chain cannot be connected to multiple widely-spaced subarrays at once, which leads to a block diagonal analog beamforming matrix. This necessitates novel design of beamforming matrices. Also, channel models with spherical wave propagation must be considered if subarrays are widely-spaced. Furthermore, the number of antenna elements per subarray, the number of subarrays, and spacing between subarrays need to be carefully designed to strike a balance between the multiplexing gain and array gain.

To overcome the channel squint effect at the THz band, THz true-time delays (TTDs) can be used to replace phase shifters in THz HBAs. TTDs allow for adjusting the phase shift relative to the carrier frequency, to generate waveforms with desired phase shifts. Moreover, to satisfy the varying dynamic channel requirements of THzCom systems, HBAs with adjustable hardware connections can be utilized. For example, a dynamic array-of-subarray (DAoS) HBA (DAoS-HBA) can be adopted, as depicted in Fig.~\ref{Fig:HBF}, and interested readers may refer to \cite{2020_Chong_DAoSA} for detailed discussion on DAoS-HBAs. Finally, a comprehensive THz HBA that incorporates wide-spaced subarrays, the DAoS, and TTDs can be designed to capture all the peculiarities of THz channels and achieves ultra-high SE and EE.

\section{Joint Integration of Other 6G-Enabling Technologies for THzCom}

To meet the wide range of QoS requirements of 6G networks, especially the connectivity, reliability, and latency requirements, e.g., ultra low-latency of less than 1-millisecond end-to-end delay and massive connectivity in the order of $10^7$ connections per $\textrm{km}^2$, several 6G-enabling technologies are anticipated to be integrated into THzCom systems to fully unlock the potential offered by the THz band \cite{2020_Mag6G_Marco_UseCasesandTechnologies,2020_WCM_THzMag_TerahertzNetworks}.
Motivated by this, in this section we discuss four 6G-enabling technologies that can be used to materialize end-to-end THzCom systems and outline the imposed challenges at physical, link, and network layers.

\subsection{Out-of-band Channel Estimation}

Accurate channel estimation is of paramount importance to effectively design and develop transmission and signal processing strategies of THzCom systems. For channel estimation in THzCom systems, on one hand, the transmitted channel state information (CSI) can be blocked  due to the high susceptibility of THz signals towards blockers. On the other hand, significant overhead can be a problem since THz channels are highly dimensional and often very sparse due to their beam representation and high reflection, diffraction, and scattering losses. These necessitate the exploration of out-of-band channel estimation for THzCom systems, i.e., estimating the CSI of THz channels using sub-6 GHz and/or mmWave band channel measurements \cite{2016_Heath_OutofBand_Journal}.

The key principle behind out-of-band channel estimation is to use spatial correlation translation techniques, which can provide the estimates of spatial correlation at the THz band from the measurements obtained from lower frequencies. Notably, the translation of correlation information from lower frequencies to the THz band poses several challenges. Specifically, the number of antennas used in antenna arrays within the same aperture at lower frequencies is less than that used at the THz band. Due to this, on one hand, a mismatch in the angles-of-arrival and the angle spread occurs between lower frequencies and THz signals. On the other hand, a large correlation matrix must be estimated from the lower frequency data obtained from fewer antennas. \textcolor{black}{These challenges open up new research avenues. For example, the THz band correlation matrix can be determined by exploiting (i) the structure of the spatial correlation matrix and interpolation/extrapolation techniques and (ii) the theoretical expressions for the THz band correlation in conjunction with the measured lower frequency correlation. We note that the latter demands the knowledge of the antenna array geometry at the receiver and the distribution of angles-of-departures.}


\subsection{Multi-connectivity}

\begin{figure}[!t]
\centering
\includegraphics[width=0.95\columnwidth]{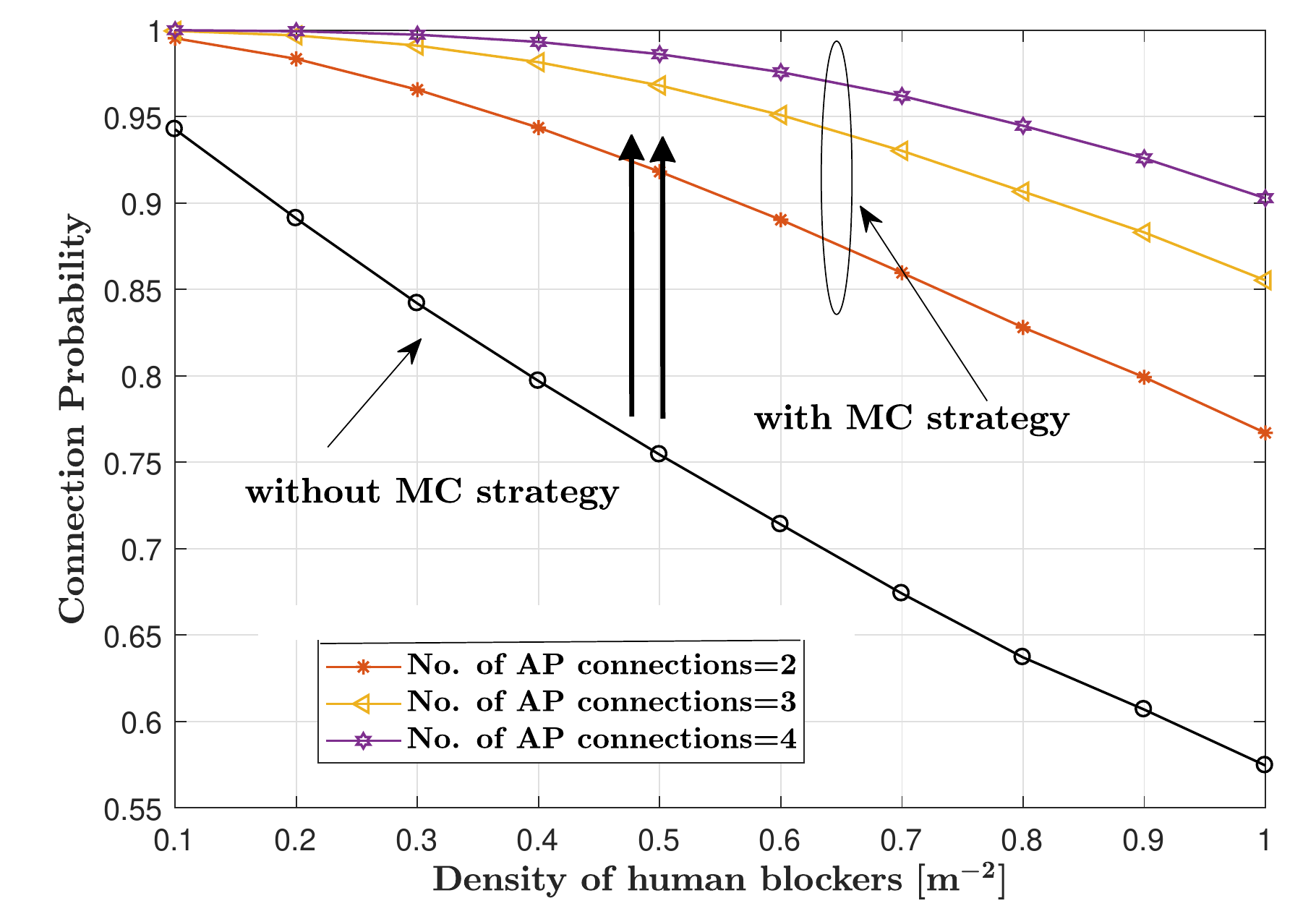} \vspace{-2mm}
\caption{Connection probability with different densities of human blockers and the number of AP connections. \textcolor{black}{In this figure, we consider the locations of APs around the user to follow a Poisson point process (PPP) with density $0.05~\textrm{m}^{-1}$. Also, we model the impenetrable dynamic human blockers by rectangular absorbing screens (commonly referred to as the double knife-edge model) with heights $1.7~\textrm{m}$  and widths $0.6~\textrm{m}$ and $0.3~\textrm{m}$, their locations by a PPP, their mobility by the random directional model, and their moving speeds by a constant velocity model~\cite{akram2020JSAC}.}}\label{Fig:MC}\vspace{-2mm}
\end{figure}

Multi-connectivity (MC) allows users to associate and communicate with multiple APs simultaneously \cite{2020_WCM_THzMag_TerahertzNetworks}. MC strategies can be effectively utilized at the THz band to overcome the reliability degradation caused by blockages. Moreover, MC and HBAs can be jointly deployed to form distributed beamforming architectures to further enhance the throughput/reliability of THzCom systems. Fig. \ref{Fig:MC} illustrates the significance of MC strategies for a user operating at the THz band, in the presence of dynamic human blockers. Through simulations, we plot the connection probability, defined as the probability that at least one associated AP has a line-of-sight (LoS) link with the user to connect and communicate with. In this figure, we consider the locations of APs that exist around the user to follow a Poisson point process with density $0.05~\textrm{m}^{-1}$, model impenetrable dynamic human blockers according to the double knife-edge model, and assume blockers' mobility following the random directional model~\cite{akram2020JSAC}. We show that the connection probability with the MC strategy is significantly higher than that without the MC strategy, especially for a high blockage density and when the user associates and communicates with a higher number of APs. This demonstrates the profound reliability improvement brought by the MC strategy.

The design of MC-enabled THzCom systems is hugely beneficial but complicated, since it requires advancements in several aspects of THzCom systems. First, the adaptation of MC strategies is likely to increase the interference in THzCom systems. \textcolor{black}{This demands new techniques to manage and mitigate interferences.} Second, MC strategies requires the cell-free architecture with centralized controlling, data processing, and combining strategies. This in turn demands high capacity backhaul connections and precise clock synchronization as well as promising packet scheduling methods for latency reduction. \textcolor{black}{For example, packet duplication-based scheduling can be explored such that packets are duplicated and the original and duplicated versions of the same packet are sent over different communication paths. Furthermore, packet splitting-based scheduling can be explored such that one packet is divided into several parts and each part of the packet is sent on different communication paths.}
\textcolor{black}{Third, dynamic AP selection strategies need to be wisely designed to improve the reliability and throughput performance in MC-enabled THzCom systems, at the cost of frequent AP switching. Given this cost, ``lightweight'' solutions in terms of software and hardware implementations are desired, aiming to provide a trade-off between improving the reliability/throughput performance and AP switching.}

\subsection{Reconfigurable Intelligent Surfaces}

Reconfigurable intelligent surfaces (RISs) can smartly steer the beams of incident radio waves in different directions by adjusting the phase shifts using a smart processor. This allows RIS to transform a LoS THz channel into a rich multipath software-defined channel with virtual LoS links. Given this benefit, RIS can be effectively utilized to mitigate the blockage vulnerability of THz signals, thereby improving the reliability of THzCom systems \cite{2018MagCombatDist,2020_WCM_THzMag_TerahertzNetworks}. We note that both passive and active RIS can be utilized at the THz band. The active RIS has the advantage of serving as distributed THz APs or signal repeaters to reflect THz signals towards specific directions by introducing arbitrary phase shifts. Moreover, RIS can be utilized to form virtual MC links when the number of APs in the system or close proximity is limited.

The research on RIS-assisted THzCom systems is in its early stages and enormous challenges lie ahead. First, given the high reflection loss at THz band, the material of reflector arrays needs to be carefully selected to support RIS at the THz band. \textcolor{black}{For example, at frequencies of 1 THz and beyond, graphene can be used for RIS, due to its tunable properties and low reflection coefficient \cite{2018MagCombatDist}.} Second, tractable models that describe the THz band RIS channel need to be developed and validated through experimental results. Third, the challenges arising during the acquisition of the CSI between the RIS and the involved AP and users need to be addressed. A viable method to overcome this challenge can be that, instead of estimating the RIS-AP and RIS-user channels separately, the concatenated AP-RIS-user channel can be estimated based on direct AP-user channel measurements and some known RIS reflection patterns.
\textcolor{black}{Fourth, to integrate RISs with the fixed wireless infrastructure, the design of a dedicated control plane, with appropriate protocols and networking procedures, is desired.}

\vspace{-2mm}
\subsection{Machine Learning Aided THzCom}

Designing effective and efficient communication paradigms in THzCom systems are far more complex compared to those at lower frequency systems, as discussed in previous sections, since channels at the THz band are known to be more uncertain than those of other frequencies. To handle this complexity while ensuring near-real-time operation, machine-learning (ML), one of the main pillars of artificial intelligence, can be used as a flexible and scalable tool to proactively optimize resources for achieving user-level and network-level performance targets \cite{2020_Mag6G_WalidSaad}.

Recently, the benefits of various ML approaches, such as supervised and unsupervised learning, reinforcement learning, and federated learning, have been studied for wireless communications systems \cite{2020VTM4}. Based on such studies, further research are worth to examine and reap the benefits of ML approaches for THzCom systems, during which several newly posed challenges need to be solved. First, data acquisition for the ML-training process can be difficult, due to the vulnerability of THz signals towards blockages and the frequency-dependent nature of THz channels. Second, processing highly non-stationary THz band training data is very challenging, since ML training algorithms that deal with non-stationary data are still in their early stages. Third, distributed network intelligence should be carefully designed based on THz device power, memory, computational limitations, security, and privacy issues related to data sharing, data transfer cost, and training complexity. By solving these challenges, ML aided THzCom systems will undoubtedly attain revolutionary leaps in channel estimation, beam tracking, blockage prediction, interference mitigation, and resource allocation, thereby bringing THzCom systems much closer to reality.

\vspace{-2mm}
\section{Conclusion}

THzCom has been envisioned as a key technology for 6G and beyond networks. Despite its promise, the research on end-to-end THzCom systems is still in its infancy. In this article, we studied pivotal areas for the development of end-to-end THzCom systems, with the emphasis on physical, link, and network layers. In particular, we examined the areas of THz spectrum management, THz band antennas and beamforming, and the integration of other 6G-enabling technologies into THzCom. This article provides a holistic view on the opportunities, key advancements, and challenges of end-to-end THzCom systems, and is expected to spur further research efforts into the use of THz band in the 6G and beyond era.

\section*{Biographies}

\noindent \footnotesize \textbf{Akram Shafie} is currently pursuing the Ph.D. degree with the School of Engineering, the Australian National University, Australia.  His current research interests include  terahertz communications, millimeter-wave communication, and machine learning in wireless
communication systems.

\vspace{2mm}

\noindent \textbf{Nan Yang} is an Associate Professor with 
the Australian National University, Australia.  His research interests include terahertz communications, ultra-reliable low latency communications, cyber-physical security, and molecular communications.
\vspace{2mm}

\noindent \textbf{Chong Han} is an Associate Professor with University of Michigan-Shanghai Jiao Tong University Joint Institute, China. His research interests include Terahertz band and millimeter-wave communication networks, and electromagnetic nanonetworks.

\vspace{2mm}

\noindent \textbf{Josep Miquel Jornet} is an Associate Professor with 
Northeastern University, USA. His research interests are in terahertz-band communications and wireless nano-bio-communication networks.

\vspace{2mm}

\noindent \textbf{Markku Juntti} is a Professor of communications engineering at the 
University of Oulu, Finland. His research interests include signal processing for wireless networks as well as communication and information theory. He is a Fellow of the IEEE.

\vspace{2mm}

\noindent \textbf{Thomas K\"urner} is a Professor of Mobile Radio Systems at the Institut f\"ur Nachrichtentechnik, Technische Universit\"at Braunschweig, Germany. His current research interests include wave propagation modeling, radio channel characterization, and radio network planning. He is a Fellow of the IEEE.

\begin{thebibliography}{10}
\providecommand{\url}[1]{#1}
\csname url@samestyle\endcsname
\providecommand{\newblock}{\relax}
\providecommand{\bibinfo}[2]{#2}
\providecommand{\BIBentrySTDinterwordspacing}{\spaceskip=0pt\relax}
\providecommand{\BIBentryALTinterwordstretchfactor}{4}
\providecommand{\BIBentryALTinterwordspacing}{\spaceskip=\fontdimen2\font plus
\BIBentryALTinterwordstretchfactor\fontdimen3\font minus
  \fontdimen4\font\relax}
\providecommand{\BIBforeignlanguage}[2]{{%
\expandafter\ifx\csname l@#1\endcsname\relax
\typeout{** WARNING: IEEEtran.bst: No hyphenation pattern has been}%
\typeout{** loaded for the language `#1'. Using the pattern for}%
\typeout{** the default language instead.}%
\else
\language=\csname l@#1\endcsname
\fi
#2}}
\providecommand{\BIBdecl}{\relax}
\BIBdecl

\bibitem{2020_Mag6G_WalidSaad}
W.~{Saad} \emph{et~al.}, ``A vision of {6G} wireless systems: Applications,
  trends, technologies, and open research problems,'' \emph{IEEE Netw.},
  vol.~34, no.~3, pp. 134--142, June 2020.

\bibitem{2020_IEEENEtworks}
K.~M.~S. Huq \emph{et~al.}, ``Terahertz-enabled wireless system for beyond-{5G}
  ultra-fast networks: A brief survey,'' \emph{IEEE Netw.}, vol.~33, no.~4, pp.
  89--95, July 2019.

\bibitem{2018MagCombatDist}
I.~F. {Akyildiz} \emph{et~al.}, ``Combating the distance problem in the
  millimeter wave and terahertz frequency bands,'' \emph{IEEE Commun. Mag.},
  vol.~56, no.~6, pp. 102--108, June 2018.

\bibitem{2019_OJCS_Survey_SalmanRef1}
H.~{Elayan} \emph{et~al.}, ``Terahertz band: The last piece of {RF} spectrum
  puzzle for communication systems,'' \emph{IEEE Open J. Commun. Soc.}, vol.~1,
  pp. 1--32, Nov. 2019.

\bibitem{2020_WCM_THzMag_Standardization}
V.~{Petrov} \emph{et~al.}, ``{IEEE} 802.15.3d: First standardization efforts
  for sub-terahertz band communications toward 6{G},'' \emph{IEEE Commun.
  Mag.}, vol.~58, no.~11, pp. 28--33, Nov. 2020.

\bibitem{2020_WCM_THzMag_TerahertzNetworks}
M.~{Polese} \emph{et~al.}, ``Toward end-to-end, full-stack {6G} terahertz
  networks,'' \emph{IEEE Commun. Mag.}, vol.~58, no.~11, pp. 48--54, Nov. 2020.

\bibitem{2020_Mag6G_Marco_UseCasesandTechnologies}
M.~{Giordani} \emph{et~al.}, ``Toward {6G} networks: Use cases and
  technologies,'' \emph{IEEE Commun. Mag.}, vol.~58, no.~3, pp. 55--61, Mar.
  2020.

\bibitem{HBM1}
Z.~{Hossain} \emph{et~al.}, ``Hierarchical bandwidth modulation for
  ultra-broadband terahertz communications,'' in \emph{Proc. IEEE Int. Conf.
  Commun. (ICC)}, Shanghai, China, May 2019, pp. 1--7.

\bibitem{HBM2}
C.~{Han} \emph{et~al.}, ``Distance-aware bandwidth-adaptive resource allocation
  for wireless systems in the terahertz band,'' \emph{IEEE Trans. THz Sci.
  Technol.}, vol.~6, no.~4, pp. 541--553, July 2016.

\bibitem{2020_Chong_TWC_DistanceAdaptiveAbsorptionPeakModulation}
W.~{Gao} \emph{et~al.}, ``Distance-adaptive absorption peak modulation for
  terahertz covert communications,'' \emph{IEEE Trans. Wireless Commun.},
  vol.~20, no.~3, pp. 2064--2077, Nov. 2020.

\bibitem{2020_Jornet_JSAC_AntennaArray}
A.~{Singh} \emph{et~al.}, ``Design and operation of a graphene-based plasmonic
  nano-antenna array for communication in the terahertz band,'' \emph{IEEE J.
  Sel. Areas Commun.}, vol.~38, no.~9, pp. 2104--2117, Sept. 2020.

\bibitem{2020_Chong_DAoSA}
L.~{Yan} \emph{et~al.}, ``A dynamic array-of-subarrays architecture and hybrid
  precoding algorithms for terahertz wireless communications,'' \emph{IEEE J.
  Sel. Areas Commun.}, vol.~38, no.~9, pp. 2041--2056, Sept. 2020.

\bibitem{2016_Heath_OutofBand_Journal}
N.~Gonzalez-Prelcic \emph{et~al.}, ``Millimeter-wave communication with
  out-of-band information,'' \emph{IEEE Commun. Mag.}, vol.~55, no.~12, pp.
  140--146, Dec. 2017.

\bibitem{akram2020JSAC}
A.~Shafie \emph{et~al.}, ``Coverage analysis for 3{D} terahertz communication
  systems,'' \emph{IEEE J. Sel. Areas Commun.}, vol.~39, no.~6, pp. 1817--1832,
  June 2021.

\bibitem{2020VTM4}
J.~{Du} \emph{et~al.}, ``Machine learning for {6G} wireless networks: Carrying
  forward enhanced bandwidth, massive access, and ultrareliable/low-latency
  service,'' \emph{IEEE Veh. Technol. Mag.}, vol.~15, no.~4, pp. 122--134,
  Sept. 2020.

\end{thebibliography}
\end{document}